\documentclass{article}

\usepackage{multicol}
\usepackage{comment}
\usepackage{amsmath}
\usepackage{enumitem}
\usepackage{url}
\usepackage{syntax}
\usepackage[dvipdfmx]{graphicx}
\usepackage{authblk}

\begin{document}

\title{Egison: Non-Linear Pattern-Matching against Non-Free Data Types}

\author{Satoshi Egi}

\maketitle

\begin{abstract}
This paper introduces the Egison programming language whose feature is strong pattern-matching facility against not only algebraic data types but also non-free data types whose data have multiple ways of representation such as sets and graphs.
Our language supports multiple occurrences of the same variables in a pattern, multiple results of pattern-matching, polymorphism of pattern-constructors and \textit{loop-patterns}, patterns that contain ``and-so-forth'' whose repeat count can be changed by the parameter.
This paper proposes the way to design expressions that have all these features and demonstrates how these features are useful to express programs concise.
Egison has already implemented in Haskell.
\end{abstract}

\section{Introduction}\label{intro}

Data types are called \textit{free} if syntactically distinct terms are unequal.
For example, lists are a free data type when we construct them with the nil and cons constructors.
When the join constructor that makes a list by appending two lists is introduced, lists are not free, but non-free, since there are multiple ways to split a list.
On the other hand, multisets and sets are always non-free data types.
This is because there are no way to provide a set of constructors to make them free since they ignore the order of the elements.

Non-free data types often appear in expressing algorithms.
Consequently, a natural way to handle them is really important.
Without it, we need to translate and regard them as a free data type whose data have a standard form when we treat them.
For example, a set would be treated as a list.
In many cases, verbose nested loops and conditional branches occur because of this translation.
For example, when we match identical pairs in a collection, we need to write nested loops and conditional branches.

We have designed a new pattern-matching system to treat non-free data types directly.
We have implemented it in our new programming language Egison~\cite{egison} using Haskell.
In this paper, we demonstrate this our new pattern-matching system and show examples to write programs utilizing this system.

\section{Demonstrations}\label{exprs}

At first, let us introduce the overview of our language.
Our language is a functional programming language with lazy evaluation strategy and has parenthesized syntax as Lisp.

\begin{grammar}
<top-expr> ::= `(define' <pat-var> <expr> `)' \hfill (top level binding)
\alt <expr>

<pat-var> :: = `$' <ident> \hfill (pattern-variable)

<expr> ::= <constant> \hfill (constant)
  \alt <ident> \hfill (variable)
  \alt `<' <Ident> <expr>* `]' \hfill (algebraic data)
  \alt `[' <expr>* `]' \hfill (tuple)
  \alt `{' <expr>* `}' \hfill (collection)
  \alt `(lambda [' <pat-var>* `]' <expr> `)' \hfill (function)
  \alt `(match-all' <expr> <expr> <match-clause> `)' \hfill (match-all expression)
  \alt `(match' <expr> <expr> `{' <match-clause>* `})' \hfill (match expression)
  \alt <matcher-expr> \hfill (matcher expression)

<match-clause> ::= `[' <pattern> <expr> `]' \hfill (match clause)

<pattern> ::= `_' \hfill (wildcard)
  \alt `$' <ident> \hfill (pattern-variable)
  \alt `,' <expr> \hfill (value-pattern)
  \alt `<' <ident> <pattern>* `>' \hfill (inductive-pattern)
  \alt `(loop' <pat-var> `[' <expr> <expr> `]' <pattern> <pattern> `)' \hfill (loop-pattern)
\end{grammar}

\textit{ident} and \textit{Ident} stand for an identifier that begin with a lowercase letter and an uppercase letter, respectively.
\texttt{match-all} and \texttt{match} expressions are syntax for pattern-matching, the core of this paper.
We explain them in detail from the next section.
\texttt{matcher} expressions are used to define how to pattern-match for each data type.
In this paper, we focus on the demonstration of our pattern-matching expressions and do not get into the mechanism behind and \texttt{matcher} expressions.

\subsection{Pattern-Matching with Backtracking}

The following is syntax of \texttt{match-all} expressions.
A \texttt{match-all} expression is composed of a \textit{target}, \textit{matcher} and \textit{match-clause}, which consists of a \textit{pattern} and \textit{body expression}.
A \texttt{match-all} expression evaluates the body of the match-clause for each pattern-matching result and returns the collection that contains all results.
A matcher specifies the way to match the target with the pattern.

\begin{grammar}
<match-all-expr> ::= `(match-all' <tgt-expr> <matcher-expr> <match-clause> `)'
\end{grammar}

Here is the first demonstration of Egison.
The only difference among the following three expressions is its matcher.
\texttt{list}, \texttt{multiset} and \texttt{set} are predefined functions in Egison core library.
They are functions that obtain a matcher and return a matcher.
For example, \texttt{(list integer)} is a matcher for a list of integers.
\texttt{(set (multiset integer))} is a matcher for a set of multisets of integers.
\texttt{integer} is a predefined matcher to pattern-match integers in Egison core library.

{\footnotesize
\begin{verbatim}
> (match-all {1 2 3} (list integer) [<cons $x $ts> [x ts]])
{[1 {2 3}]}
> (match-all {1 2 3} (multiset integer) [<cons $x $ts> [x ts]])
{[1 {2 3}] [2 {1 3}] [3 {1 2}]}
> (match-all {1 2 3} (set integer) [<cons $x $ts> [x ts]])
{[1 {1 2 3}] [2 {1 2 3}] [3 {1 2 3}]}
\end{verbatim}
}

In the above expressions, \texttt{<cons \$x \$ts>} is a pattern.
\texttt{cons} is a pattern-constructor.
The name of a pattern-constructor starts with lowercase.
It divides a collection into a head element and the rest.
The meaning of a head differs for each matcher.
`\texttt{\$x}' and `\texttt{\$ts}' are called \textit{pattern-variables}.
We can access the result of pattern-matching referring to them.

The characteristic of our pattern-matching expression is it takes a matcher.
It realizes polymorphism of pattern-constructors and enables us to use the same pattern-constructors for similar data types.
We specifies a matcher in pattern-match expressions, because data of non-free data types such as a collection can be pattern-matched as different data types in many places of programs.

We introduce other pattern-constructors \texttt{nil} and \texttt{join}.
The \texttt{nil} pattern-constructor takes no arguments and matches when the target is an empty collection.
The \texttt{join} pattern-constructor takes two arguments and divides a collection into two collections.
The following is a demonstration of \texttt{join}.

{\footnotesize
\begin{verbatim}
> (match-all {1 2 3} (list integer) [<join $xs $ys> [xs ys]])
{[{} {1 2 3}] [{1} {2 3}] [{1 2} {3}] [{1 2 3} {}]}
\end{verbatim}
}

Finally, we can handle pattern-matching that has even infinite results.

{\footnotesize
\begin{verbatim}
> (take 8 (match-all nats (set integer) [<cons $m <cons $n _>> [m n]]))
{[1 1] [1 2] [2 1] [1 3] [2 2] [3 1] [1 4] [2 3]}
\end{verbatim}
}

\texttt{take} is a function that obtains a number \textit{n} and a collection \textit{xs} and returns the first \textit{n} elements of \textit{xs}.
\texttt{nats} is an infinite list that contains all natural numbers.
`\texttt{_}' is an wildcard and matches with any object.
Our pattern-matching system guarantees to enumerate all successful matching results.
The idea of the traverse strategy in the pattern-matching process is similar with the idea of this paper~\cite{spivey2000functional}.
In brief, we adopt breadth-first search for the traverse strategy.

\subsection{Non-Linear Pattern-Matching}

Non-linear patterns are patterns that allow multiple occurrences of same variables in a pattern.
The following is an example of a non-linear pattern.
The output of this example is the collection of numbers from which three number sequence starts.

{\footnotesize
\begin{verbatim}
> (match-all {1 5 6 2 4} (multiset integer)
    [<cons $n <cons ,(+ n 1) <cons ,(+ n 2) _>>> n])
{4}
\end{verbatim}
}

A pattern is examined from left to right in order, and the binding to a pattern-variable can be referred to in its right side of the pattern.
In this example, at first, the pattern-variable `\texttt{\$n}' is bound to any element of the collection since the matcher is \texttt{(multiset integer)}.
After that, the value-pattern `\texttt{,(+ n 1)}' and `\texttt{,(+ n 2)}' are examined.
A value-pattern begins with `\texttt{,}'.
The expression following `\texttt{,}' can be any kind of expressions.
In this case, the value-patterns match with a target if the target object is equal with the content of the pattern.
Therefore, after successful pattern-matching, `\texttt{\$n}' is bound to an element from which three number sequence starts.

How to handle value-patterns is defined in matchers, and then varies by matchers.
For example, the way to check equality for integers is defined in the \texttt{integer} matcher.
It realizes polymorphism of value-patterns.
For example, the way to check equality for lists and multisets are different, but we can use value-patterns for both of them as we can use same pattern-constructors for them.
The advantage of the value-pattern notation over guard is it enables us to read a pattern with the same order of the execution process of pattern-matching.

Let us show another demonstration of non-linear patterns.
It enumerates all twin primes by pattern-matching against the infinite list of prime numbers.

{\footnotesize
\begin{verbatim}
> (define $twin-primes
    (match-all primes (list integer)
      [<join _ <cons $p <cons ,(+ p 2) _>>> [p (+ p 2)]]))
> (take 6 twin-primes)
{[3 5] [5 7] [11 13] [17 19] [29 31] [41 43]}
\end{verbatim}
}

We can write pattern-matching against nested non-free data types such as a list of multisets or a set of sets as follow.

{\footnotesize
\begin{verbatim}
> (match-all {{1 2 3 4 5} {4 5 1} {6 1 7 4}} (list (multiset integer))
   [<cons <cons $n _> <cons <cons ,n _> <cons <cons ,n _> <nil>>>> n])
{1 4}
\end{verbatim}
}

Our language has \texttt{match} expressions as other functional languages.
A \texttt{match} expression takes multiple match-clauses and tries pattern-matching for each pattern from the head of match-clauses.
A \texttt{match} expression is useful to express conditional branches.

\begin{grammar}
<match-expr> ::= `(match' <tgt-expr> <matcher-expr> `\{' <match-clause>* `\})' 
\end{grammar}

The following is a demonstration that determines poker-hands.
Note that all poker-hands are represented in a single pattern.
The \texttt{card} matcher is defined in the same way with algebraic data types of the existing functional languages.

\begin{multicols}{2}
{\scriptsize
\begin{verbatim}
(define $poker-hands
  (lambda [$cs]
    (match cs (multiset card)
      {[<cons <card $s $n>
         <cons <card ,s ,(- n 1)>
          <cons <card ,s ,(- n 2)>
           <cons <card ,s ,(- n 3)>
            <cons <card ,s ,(- n 4)>
             <nil>>>>>>
        <Straight-Flush>]
       [<cons <card _ $n>
         <cons <card _ ,n>
          <cons <card _ ,n>
            <cons <card _ ,n>
              <cons _
                <nil>>>>>>
        <Four-of-Kind>]
       [<cons <card _ $m>
         <cons <card _ ,m>
          <cons <card _ ,m>
           <cons <card _ $n>
            <cons <card _ ,n>
              <nil>>>>>>
        <Full-House>]
       [<cons <card $s _>
         <cons <card ,s _>
           <cons <card ,s _>
             <cons <card ,s _>
               <cons <card ,s _>
                 <nil>>>>>>
        <Flush>]
       [<cons <card _ $n>
         <cons <card _ ,(- n 1)>
          <cons <card _ ,(- n 2)>
           <cons <card _ ,(- n 3)>
            <cons <card _ ,(- n 4)>
             <nil>>>>>>
        <Straight>]
       [<cons <card _ $n>
         <cons <card _ ,n>
          <cons <card _ ,n>
           <cons _
            <cons _
             <nil>>>>>>
        <Three-of-Kind>]
       [<cons <card _ $m>
         <cons <card _ ,m>
          <cons <card _ $n>
            <cons <card _ ,n>
             <cons _
               <nil>>>>>>
        <Two-Pair>]
       [<cons <card _ $n>
         <cons <card _ ,n>
          <cons _
           <cons _
            <cons _
             <nil>>>>>>
        <One-Pair>]
       [<cons _
         <cons _
          <cons _
           <cons _
            <cons _
             <nil>>>>>>
        <Nothing>]})))
\end{verbatim}
  }
\end{multicols}

\subsection{Loop Patterns}

Let us consider a function \texttt{comb2} that takes a collection returns the 2-combinations of the elements.
The function is written using pattern-matching as follow.
\texttt{something} is a only built-in matcher and it can be used only for pattern-matching with a wildcard or a pattern variable.

{\footnotesize
\begin{verbatim}
> (define $comb2 (lambda [$xs]
    (match-all xs (list something)
     [<join _ <cons $a_1 <join _ <cons $a_2 _>>>> {a_1 a_2}])))
> (comb2 {1 2 3 4})
{{1 2} {1 3} {2 3} {1 4} {2 4} {3 4}}
\end{verbatim}
}

Now, we explain \textit{indexed-variables}.
A variable whose name is followed by `_' and an expression is an indexed-variable.
The expression after `_' must be evaluated to a natural number and is called an \textit{index}.
We can append as many indexes as we want.

Next, let us consider  \texttt{comb3}, \texttt{comb4}, \texttt{comb5}, and so on.
Patterns in these \texttt{combX} have the same form, \texttt{<join \_ <cons \$a\_1 <join \_ <cons \$a\_2 ... \_>...>>>}.
It seems to be possible to generalize them.
A \textit{loop-pattern} is for such a purpose.
A loop-pattern has the following syntax.

\begin{grammar}
<loop-pat> ::= `(loop' <pat-var> `[' <idx> <last-num> `]' <rep-pat> <tail-pat> `)'
\end{grammar}

The arguments of a loop-pattern respectively represent an index-variable, a range of the index, a pattern repeated, and a pattern at the end.
A range of index is represented with a tuple that consists of a number where index starts, which is called the \textit{current index} and a number where index ends, which is called the \textit{last index}.
We can define \texttt{comb} which handles general n-combinations of the elements as follow.

{\footnotesize
\begin{verbatim}
> (define $comb (lambda [$xs $n]
    (match-all xs (list something)
     [(loop $i [1 n] <join _ <cons $a_i ...>> _)
      (map (lambda [$i] a_i) (take n nats))])))
> (comb {1 2 3 4} 2)
{{1 2} {1 3} {2 3} {1 4} {2 4} {3 4}}
> (comb {1 2 3 4} 3)
{{1 2 3} {1 2 4} {1 3 4} {2 3 4}}
\end{verbatim}
}

We explain how above code is interpreted when \texttt{n} is \texttt{2}.
A loop-pattern is \texttt{(loop \$i [1 2] <join \_ <cons \$a\_i ...>> \_)}.
When the interpreter meets a loop-pattern and the current index is not greater than the last index, a loop-pattern returns the third argument replacing `\texttt{...}' with the loop-pattern itself.
Therefore, the above example is evaluated to \texttt{<join \_ <cons \$a\_1 (loop \$i [2 2] <join \_ <cons \$a\_i ...>> \_)>>}.
Note that in the evaluation, the index-variable \texttt{i} is replaced with the first index, \texttt{1} in the example.
Moreover, the current index of the extended loop-pattern proceeds to `2` from `1`.
That is, \texttt{[1 2]} is replaced with \texttt{[2 2]}.
Repeating this evaluation again, we reach
\texttt{<join \_ <cons \$a\_1 <join \_ <cons \$a\_2 (loop \$i [3 2] <join \_ <cons \$a\_i ...>> \_)>>>>}.
When the current index is greater than the last index, a loop-pattern returns the fourth argument.
So in the case, the loop-pattern is replaced with `\texttt{\_}'.
Then, we get \texttt{<join \_ <cons \$a\_1 <join \_ <cons \$a\_2 \_>>>>}.
It is the same pattern we used in \texttt{comb2}.

In the above, we omit explanation of restriction about the place of `\texttt{...}'.
`\texttt{...}' must be placed at the end of the second argument.
For example, \texttt{<cons ... <nil>>} is prohibited.
This restriction decides which loop-patterns a given `\texttt{...}' belongs to, and then allows us to write nested loop-patterns.

If we use loop-patterns in a pattern, the count of the pattern-variables in the pattern can change by the parameter.
It is the reason why we introduced indexed-variables.

\section{Pattern-Matching-Oriented Programming Style}\label{style}

We can redefine the well-known list library's functions with pattern-matching.
\texttt{eq} is a predefined matcher for data types on whom equality is syntactically defined.
When the \texttt{eq} matcher is used, equality is checked for value-patterns.

{\footnotesize
\begin{verbatim}
(define $map (lambda [$xs $fn] (match-all xs (list something)
  [<join _ <cons $x _>> (fn x)])))

(define $member? (lambda [$x $xs] (match xs (multiset eq)
  {[<cons ,x _> <True>] [_ <False>]})))

(define $delete (lambda [$x $xs] (match xs (list eq)
  {[<join $hs <cons ,x $ts>> (append hs ts)] [_ xs]})))

(define $take (lambda [$n $xs] (match xs (list something)
  {[(loop $i [1 n] <cons $a_i ...> _)
    (map (lambda [$i] a_i) (take n nats))]})))
\end{verbatim}
}

\section{Related Work}\label{related}

In this section, we introduce existing studies in the field of pattern-matching.

McBride's symbol manipulation system~\cite{mcbride1970symbol} may be the first non-linear pattern-matching system.
In his paper, there are several demonstrations to process math expressions that show the expressive power of non-linear patterns.
However, McBride's approach does not support pattern-matching with backtracking and only supports pattern-matching against a list as a collection.

Wadler's views~\cite{wadler1987views} provide the way to decompose data with multiple representations, by declaring transformation between each representations.
For example, we can intuitively handle complex numbers that have cartesian and polar representation.
Data are automatically transformed in the matching process.
However, they treat neither multiple results of pattern-matching nor non-linear patterns.

Active patterns~\cite{erwig1996active} provide a way to decompose non-free data.
For example, we can implement the \texttt{cons} pattern constructor for multiset with active patterns.
The limitation of active patterns is that it does not support backtracking.
Therefore, for example, we cannot write a pattern that matches identical pairs in a collection.

First class patterns~\cite{tullsen2000first} propose a sophisticated system that treats patterns as first class objects.
First class patterns can deal with pattern-matching that has multiple results.
However pattern-matching with this proposal also has limitation that it does not support non-linear pattern-matching.

Functional logic programming~\cite{AntoyHanus05LOPSTR} is another approach.
It applies unification of logic programming to pattern-matching.
So it can handle both of non-linear patterns and backtracking.
The progress by our proposal from this work is polymorphism of pattern-constructors by matchers, the support for infinite results of pattern-matching and loop-patterns.
Actually, there is also a difference in the mechanism behind and the way to define matchers between this work and our proposal though we do not focus on it in this paper.

\section{Conclusion}\label{conclusion}

The contribution of our proposal is an invention of a pattern-matching system with \textbf{all of the following features} in functional programming.
Additionally, we set a example to write programs utilizing these features.
We contribute to programming language community by extending the area that we can write a program in more concise way.

\begin{description}
  \item \textbf{Non-linear patterns}

    We can handle multiple occurrences of same variables in a pattern.
    Non-linear patterns are represented with value-patterns that allow us to write expressions in a pattern.

  \item \textbf{Multiple pattern-matching results}

    We can handle pattern-matching that has multiple and even infinite results.
    This feature is necessary for pattern-matching against data types whose data have multiple way of decomposition.

  \item \textbf{Polymorphism of pattern-constructors}

    We can use the same pattern-constructors for similar data types.
    This feature reduces the number of names of pattern-constructors to remember.
    Just with \texttt{nil}, \texttt{cons} and \texttt{join}, we can express most of patterns against collections.

  \item \textbf{Loop patterns}

    Loop-patterns enables to express patterns that the count of the pattern-variables in the pattern can change by the parameter.

\end{description}

We believe direct and concise representation of algorithms promotes us to implement really new things that we have never imagined to implement.
We hope our work will make breakthroughs in various fields.

\bibliographystyle{abbrvnat}
\bibliography{egison}

\end{document}